\documentclass[prl,12pt]{revtex4}
\pdfoutput=1
\usepackage{amsmath}
\usepackage{graphicx}
\usepackage{subfigure}
\usepackage{bm}
\usepackage{slashed}

\def\d{\mathrm{d}}

\begin{document}
\title{Predicting the isospin asymmetry in $B\rightarrow K^* \gamma$ using holographic AdS/QCD Distribution Amplitudes for the $K^*$}
\author{M. Ahmady \\
Department of Physics, Mount Allison University \\
Sackville, N-B, Canada\\
\email{} }
\author{R. Sandapen \\
D\'epartement de Physique et d'Astronomie, Universit\'e de Moncton, \\
Moncton, N-B. E1A 3E9, Canada.\\
\& \\
Department of Physics, Mount Allison University \\
Sackville, N-B, Canada\\
\email{ruben.sandapen@umoncton.ca} }
\preprint{}
\begin{abstract}
We predict the isospin asymmetry as well as the branching ratio for the decay $B \to K^* \gamma$ within QCD factorization using new anti-de Sitter/Quantum Chromodynamics (AdS/QCD) holographic Distribution Amplitudes (DAs) for the $K^*$ meson. Our prediction for the branching ratio agrees with that obtained using standard QCD Sum Rules (SR) DAs and with experiment. More interestingly, 
our prediction for the isospin asymmetry using the AdS/QCD DA  does not suffer from the end-point divergence encountered when using the corresponding SR DA.  We predict an isospin asymmetry of $3.2\%$ in agreement with the most recent average measured value of $(5.2 \pm 2.6)\%$ quoted by the Particle Data Group. 
\end{abstract}

\maketitle

\section{Introduction}
The rare decay $B \to K^* \gamma$ is the dominant mode of exclusive radiative decays of the $B \to V \gamma$ where $V$ is a vector meson. It was first observed by the CLEO collaboration in 1993 \cite{PhysRevLett.71.674} and since then it has been measured with increasing precision by the BaBar\cite{PhysRevLett.103.211802} , Belle \cite{PhysRevD.69.112001} and CLEO \cite{PhysRevLett.84.5283} collaborations. The most recent data for the branching ratios of the decay $B^{\circ} \to K^{*0} \gamma$ and $B^+ \to K^{*+} \gamma$ are given in Table \ref{tab:BRdata}. As can be seen, all three experiments report a slightly higher branching ratio for $B^{\circ} \to K^{*0} \gamma$ and this indicates a non-zero isospin asymmetry defined 
as
\begin{equation}
\Delta_{0-}=\frac{\Gamma (\bar{B}^{\circ} \to \bar{K^*} \gamma) - \Gamma(B^{-} \to K^{*-} \gamma)}{\Gamma (\bar{B}^{\circ} \to \bar{K^*} \gamma) + \Gamma(B^{-} \to K^{*-} \gamma)} \;.
\label{Isospin}
\end{equation}
The most recent isospin asymmetry measurements reported by BaBar and Belle are shown in the last row of table \ref{tab:BRdata}.  Note that $\Delta_{0+}=\Delta_{0-}$ to within $2\%$ which is the maximum measured CP asymmetry for this decay.

\begin{table}[h]
\begin{center}
\[
\begin{array}
[c]{|c|c|c|c|c|}\hline
\mbox{Branching ratio}& \mbox{BABAR}  &\mbox{BELLE} &\mbox{CLEO}  & \mbox{PDG}  \\ \hline

{\cal B}(B^0\to K^{*0}\gamma )\times 10^6&44.7\pm 1.0 \pm 1.6 &45.5^{+7.2}_{-6.8}\pm 3.4 &40.1\pm 2.1\pm 1.7 & 43.3 \pm 1.5\\ \hline
{\cal B}(B^+\to K^{*+} \gamma )\times 10^6 & 42.2\pm 1.4 \pm 1.6 & 42.5\pm 3.1\pm 2.4 & 37.6^{+8.9}_{-8.3}\pm 2.8  & 42.1\pm 1.8 \\ \hline
\Delta_{0-} &6.6\pm 2.1\pm 2.2& 1.2\pm 4.4\pm 2.6 & &5.2 \pm 2.6 \\ \hline
\end{array}
\]
\end{center}
\caption {The measured branching ratios and isospin asymmetry.}
\label{tab:BRdata}
\end{table}

In general,  radiative $B$ decays to vector mesons are of considerable interest because they proceed via Flavor Changing Neutral Currents  (FCNC) which are heavily suppressed at tree level in the Standard Model (SM) and are thus likely to be enhanced by  New Physics (NP) \cite{Ahmady:2001qh}. Such exclusive decays are also relatively clean to investigate experimentally especially in a hadronic environment like the LHC. On the other hand, the theory of  exclusive decays  is complicated by their sensitivity to non-perturbative physics. Nevertheless,  it is very important to have reliable SM predictions for these decays in order to detect any NP signals.

The standard theoretical framework for computing exclusive radiative $B$ decays is QCD factorization (QCDF) \cite{Bosch:2001gv}.   QCDF is the statement that to leading power accuracy in the heavy quark limit, the matrix element of the effective weak Hamiltonian operators factorizes into perturbatively calculable kernels and non-perturbative but universal quantities namely the $B \to V$ transition form factor  and the leading twist DAs of the B and vector mesons.  In a standard notation, these matrix elements are written as  \cite{Bosch:2001gv,Ball:2006eu}

\begin{equation}
\langle V(P,e_T) \gamma (q,\epsilon ) | Q_i | \bar{B} \rangle = [ F^{B\rightarrow V} T_i^I  + \int_0^1 \d \zeta\; \d z \; \Phi_B(\zeta) T_i^{II}(\zeta,z)  \phi^{\perp}_{V} (z)] \cdot \epsilon   + \mathcal{O}(\Lambda_{\mbox{\tiny{QCD}}}/m_b)  \;.
\label{factorization}
\end{equation}
The first term is simply the product of perturbatively calculable quantities $T_i^I$  with the non-perturbative  transition form factor $F^{B\rightarrow V}$. The second term is a convolution of the perturbatively computable kernels $T_i^{II}$ with the non-perturbative DA of the $B$ meson, $\Phi_B(\zeta)$, and the DA of vector meson, $\phi_{V}^{\perp}(z)$, where $z$ is the fraction of the meson light-front momentum carried by the quark. Formally, the second term is a correction of order $\alpha_s$ to the first term. Traditionnally, the form factor and DAs are obtained from  QCD Sum Rules or lattice QCD.

To leading power accuracy in the heavy quark mass, the decay amplitude for $B \to K^* \gamma$ is given by  \cite{Bosch:2001gv}\footnote{Neglecting the CKM supressed contribution.}
\begin{equation}
\mathcal{A}_{\mbox{\tiny{Leading}}} (B \to K^* \gamma)= \frac{G_F}{\sqrt{2}} V_{cs}^* V_{cb}  a_7^c \langle K^*(P,e_T) \gamma (q,\epsilon ) | Q_7 | \bar{B} \rangle 
\label{leading-amp}
\end{equation}
where at next-to-leading order (NLO) in the strong coupling, $a_7^c$ is given by \cite{Bosch:2001gv}
\begin{eqnarray}
a_7^c &=&C_7+\frac{\alpha_s(\mu )C_F}{4\pi} [C_1(\mu)G_1(s_c)+C_8(\mu )G_8] \nonumber \\
&+&\frac{\alpha_s(\mu_{h})C_F}{4\pi}[C_1(\mu_{h} )H_1(s_c,\mu_{h})+C_8(\mu_{h} )H_8(\mu_{h})] \;.
\label{a7}
\end{eqnarray} 
In Eq. \eqref{a7}, the strong coupling $\alpha_s$, the Wilson coeffecients $C_{1,8}$ and the functions $H_{1,8}$ are evaluated at two different scales: a hard scale $\mu=m_b$ and a hadronic scale $\mu_{h}=\sqrt{\Lambda_{\tiny{\mbox{QCD}}} \mu} \approx 2~\mbox{GeV}$. The function $H_1$ also depends on $s_c=(m_c/m_b)^2$ where $m_c$ is the charm quark mass in loops contributing at  NLO accuracy in $\alpha_s$. The explicit expressions for  the hard scattering functions $G_1(s_c)$ and $G_8$ can be found in Ref. \cite{Bosch:2001gv}.  Here, it suffices to specify the integrals
\begin{equation}
H_1(s_c,\mu_h) =-\left( \frac{2 \pi^2 f_B f_{K^*}^{\perp}(\mu_h)}{3N_c M_B^2}\right) \left( \frac{M_B}{\lambda_B}\right)\int_0^1 \d z \; h(s_c,\bar{z}) \phi_{K^*}^{\perp} (z,\mu_h) 
\label{I1}
\end{equation}
and
\begin{equation}
H_8(\mu_h) = \left( \frac{4 \pi^2 f_B f_{K^*}^{\perp}(\mu_h)}{3N_c F^{B\to K^*} M_B^2}\right) \left( \frac{M_B}{\lambda_B}\right) \int_0^1 \d z \; \frac{\phi_{K^*}^{\perp}(z,\mu_h)} {z} 
\label{I2}
\end{equation}
which depend on the twist-$2$ DA of the transversely polarised $K^*$ meson, $\phi_{K^*}^{\perp}$. In Eq. \eqref{I1} and \eqref{I2},  $f_{K^*}^{\perp}$ is the scale-dependent decay constant of the $K^*$ meson, $f_B$ is the decay constant  of the $B$ meson and $ h(s_c,\bar{z})$, with $\bar{z}=1-z$, is a hard scattering kernel given explicitly in Ref. \cite{Bosch:2001gv}. The quantity $M_B/\lambda_B$, where $M_B$ is the mass of the $B$ meson and $\lambda_B \sim \Lambda_{\mbox{\tiny{QCD}}}$, parametrizes the inverse moment of the $B$ meson DA.  

The branching ratio corresponding to the leading power amplitude is then given by 
 \begin{equation}
\label{BR}
 {\cal BR}(B \to K^* \gamma )=\frac{\tau_{B}}{16\pi M_{B}^3}\left (1-\left(\frac{M_{K^*}}{M_{B}}\right)^2\right )
|\mathcal {A}_{\mbox{\tiny{Leading}}}  (B \to K^* \gamma )|^2  
\end{equation}
where $M_{K^*}$ is the mass of the $K^*$ meson and $\tau_B$ is the measured lifetime  of the $B$ meson. 

The predictive power of QCDF is therefore limited by two sources of uncertainty: firstly by the uncertainties associated with the non-perturbative quantities (form factor, decay constants and DAs) which we shall refer to as hadronic uncertainties and secondly  by power corrections to the leading contribution given by Eq. \eqref{factorization}. The computation of power corrections is problematic because it involves convolution integrals that do not always converge \cite{Pecjak:2008gv,Antonelli:2009ws}. In a recent paper \cite{Ahmady:2012dy}, we have investigated such power corrections in the decay $\bar{B}^\circ \to \rho^{\circ} \gamma$ namely those generated by annihilation diagrams. Two of the four annihilations contributions we considered  depend  upon convolution integrals involving the vector twist-$3$ DA of the vector meson. These integrals  diverge at the end-points when using the standard Sum Rules twist-$3$ DA. We found that these divergences are avoided when using alternative AdS/QCD twist-$3$ DA for the $\rho$ meson. Nevertheless, we confirmed that the annihilation power corrections to the leading amplitude are numerically small so that the end-point divergence problem has no practical consequences when computing the branching ratio for this decay. We expect this to be also the case for the decay $B \to K^* \gamma$. The main uncertainties in computing the branching ratio are therefore the hadronic uncertainties. 

On the other hand, the isospin asymmetry given by Eq. \eqref{Isospin} is less sensitive to the hadronic uncertainties since it depends on the ratio of decay rates or equivalently on the ratio of branching ratios.  However,  in computing this observable for $B \to K^* \gamma$,  the end-point divergence problem cannot be ignored. This is because the isospin asymmetry in $B \to K^* \gamma$ vanishes to leading power accuracy and any deviation from zero is due to power-supressed contributions.  These can be parametrized as $\mathcal{A}_{q}=b_q \mathcal{A}_{\mbox{\tiny{leading}}}$ where $q$ is the flavor of the spectator antiquark in the $B$ meson \cite{Kagan:2001zk}.  To leading order in small quantities, the isospin asymmetry is then given by \cite{Kagan:2001zk}
\begin{equation}
\Delta_{0-} = \Re \mbox{e}(b_d -b_u)
\label{isospin}
\end{equation}
with 
\begin{equation}
b_q=\frac{12 \pi^2 f_B Q_q}{m_b F^{B \to K^*} a_7^c} \left( \frac{f_{K^*}^{\perp}}{m_b} K_1 + \frac{f_{K^*} M_{K^*}}{6 \lambda_B M_B} K_2 \right)
\end{equation}
where $K_1$ and $K_2$ are dimensionless coefficients given explicitly in \cite{Kagan:2001zk}. They depend on four convolution integrals namely
\begin{equation}
F_{\perp} (\mu_h)=\int_0^1 \d z \;  \frac{\phi^{\perp}_{K^*}(z,\mu_h)}{3(1-z)}
\label{Fperp}
\end{equation}
\begin{equation}
G_{\perp}(s_c,\mu_h)=\int_0^1 \d z \; \frac{\phi^{\perp}_{K^*}(z, \mu_h)}{3(1-z)} G(s_c,\bar{z})
\label{Gperp}
\end{equation}
\begin{equation}
X_{\perp}( \mu_h) = \int_0^1 \d z \; \phi_{K^*}^{\perp}(z, \mu_h) \left( \frac{1 + \bar{z}}{3 \bar{z}^2} \right)
\label{Xperp}
\end{equation}
and
\begin{equation}
H_{\perp}(s_c, \mu_h) = \int_0^1 \d z \; \left(g^{\perp(v)}_{K^*} (z,\mu_h) - \frac{1}{4}\frac{\d g_{K^*}^{\perp (a)}}{\d z} (z,\mu_h)\right)G(s_c,\bar{z})
\label{Hperp}
\end{equation}
where $G(s_c,\bar{z})$ is the penguin function \cite{Kagan:2001zk}.  The first three integrals $F_{\perp}$, $G_{\perp}$ and $X_{\perp}$ depend on  the twist-$2$ DA  while $H_{\perp}$ depends  on the twist-$3$ DAs.  It turns out that $X_\perp(\mu)$ diverges with the standard SR twist-$2$ DA  \cite{Kagan:2001zk}. 

 This isospin asymmetry was first computed in Ref. \cite{Kagan:2001zk} using Sum Rules DAs evaluated at a scale $\mu_h=\sqrt{5}$ GeV. The diverging integral $X_{\perp}$ was regulated using a cut-off, thus introducing an additional uncertainty in the theoretical prediction. In Ref. \cite{Ball:2006eu}, the contribution of the divergent integral was neglected while other contributions beyond QCDF, namely long distance photon emission and gluon emission from quark loops, were taken into account. 

Our goal in this paper is to compute the isospin asymmetry given by Eq. \eqref{isospin} as well as the branching ratio given by Eq. \eqref{BR} using holographic AdS/QCD DAs for the transversely polarized $K^*$ meson. In doing so, we shall show that the end-point divergence in $X_\perp$ can be avoided and that we predict an isospin asymmetry that is consistent with experiment. Moreover, we shall see that our AdS/QCD prediction for the branching ratio at leading power accuracy agrees with the Sum Rules prediction  and with experiment.

We now turn to the derivation of the holographic AdS/QCD DAs of the $K^*$ meson. They are obtained using an AdS/QCD holographic light-front wavefunction  \cite{deTeramond:2008ht} for the $K^*$ meson. Our derivation is a generalisation of our earlier derivation \cite{Ahmady:2012dy}  for the AdS/QCD DAs of the $\rho$ meson.  We now account for unequal quark masses and thus for the resulting $SU(3)$ flavor symmetry breaking effects.

\section{Holographic AdS/QCD  Distribution Amplitudes}

The AdS/QCD holographic wavefunction \cite{deTeramond:2008ht,Brodsky:2008kp} for a
ground state vector meson  in which the quark of mass $m_q$ carries
a fraction $z$ of the meson light-front momentum\footnote{The
antiquark of mass $m_{\bar{q}}$ then carries $(1-z)$ of the meson
light-front momentum.}, can be written as \cite{Vega:2009zb}
\begin{equation}
\phi_{\lambda} (z,\zeta)= \mathcal{N}_{\lambda}
\frac{\kappa}{\sqrt{\pi}}\sqrt{z(1-z)} \exp
\left(-\frac{\kappa^2 \zeta^2}{2}\right)
\exp\left(-\frac{(1-z)m_{q}^2 + zm_{\bar{q}}^2}{2\kappa^2 z (1-z)}
\right)~, \label{AdS-QCD-wfn}
\end{equation}
where $\zeta=\sqrt{z(1-z)} r$ with $r$ being the transverse separation between the
quark and antiquark. This wavefunction is obtained by solving the AdS/QCD holographic light-front Schroedinger equation \cite{deTeramond:2008ht} for mesons where the interacting potential in four dimensional 
physical spacetime is determined by  the dilaton background field that breaks conformal invariance in five dimensional AdS space. Theoretical and phenomenological considerations constraint the form of the dilaton field 
to be quadratic \cite{Brodsky:2013npa} . In that case,  the parameter $\kappa$ is fixed by the meson mass: $\kappa=M_{V}/\sqrt{2}$. 
Note that we allow the normalization constant $\mathcal{N}_{\lambda}$ to depend on the polarization $\lambda=L,T$ of the vector meson
\cite{Forshaw:2012im}. For the $K^*$ vector meson, $M_V=M_{K^*}$, $q=s$ and $\bar{q}=\bar{u}$ or $\bar{d}$.  

The AdS/QCD wavefunction of the $K^*$ vector meson can thus be written as
\begin{equation}
\phi_{K^*}^{\lambda} (z,\zeta)= \mathcal{N}_{\lambda}
\frac{\kappa}{\sqrt{\pi}}\sqrt{z(1-z)} \exp
\left(-\frac{\kappa^2 \zeta^2}{2}\right)
\exp\left \{-\left[\frac{m_s^2-z(m_s^2-m^2_{\bar{q}})}{2\kappa^2 z (1-z)} \right]
\right \} \label{AdS-QCD-wfn-K*}
\end{equation}
with $\kappa=0.63$ GeV and where we have made explicit the $SU(3)$ flavor symmetry breaking correction proportional to $(m^2_s-m^2_{\bar{q}})$ in the second term in the last exponential. 

The meson light-front wavefunctions can be written in terms of the
AdS/QCD wavefunction.  In momentum space \cite{Forshaw:2011yj}
\begin{equation}
\Psi^{K^*,\lambda}_{h,\bar{h}}(z,\mathbf{k})=\sqrt{\frac{N_c}{4\pi}}
S_{h,\bar{h}}^{K^*,\lambda}(z,\mathbf{k}) \phi_{K^*}^{\lambda}(z,\mathbf{k})
\label{spinor-scalar-k}
\end{equation}
where
\begin{equation}
 S_{h,\bar{h}}^{K^*,\lambda}(z,\mathbf{k})=\frac{\bar{u}_{h} (zP^+ ,
-\mathbf{k} )}{\sqrt{z}} e^{\lambda}.\gamma
\frac{v_{\bar{h}}((1-z)P^+,\mathbf{k})}{\sqrt{(1-z)}}
\label{Slambda}
\end{equation}
and $\phi_{K^*}^{\lambda}(z,\mathbf{k})$ is the two dimensional Fourier
transform of the AdS/QCD wavefunction given by Eq. \eqref{AdS-QCD-wfn}. Note that in Eq. \eqref{Slambda}, $h$ is the helicity of quark and $\bar{h}$ is the helicity of the antiquark.
The normalization $\mathcal{N}_{\lambda}$ of the AdS/QCD wavefunction is fixed by imposing
that \cite{Forshaw:2003ki,Forshaw:2012im}
\begin{equation}
\sum_{h,\bar{h}} \int \frac{\d^2 \mathbf{k}}{(2\pi)^2} |\Psi_{h,\bar{h}}^{K^*,\lambda}(z,\mathbf{k})|^2 =1 \;.
\end{equation}
Choosing the longitudinal and transverse polarization vectors as
\begin{equation}
e^{L}=\left(\frac{P^{+}}{M_{K^*}},-\frac{M_{K^*}}{P^{+}},0_{\perp}
\right) ~\hspace{1cm}~\mbox{and}\hspace{1cm}
e^{T}=\frac{1}{\sqrt{2}}\left(0,0, 1, \pm i \right)
\end{equation}
where $P^+$ is the ``plus'' component of the $4$-momentum of the $K^*$ meson given by
\begin{equation}
P^{\mu}=\left(P^{+},\frac{M_{K^*}^2}{P^{+}},
0_{\perp} \right)
\end{equation}
and using the light-front spinors of reference \cite{Lepage:1980fj}, we find that the spinor wavefunctions are given by 
\begin{equation}
S_{h,\bar{h}}^{K^*,L}(z,\mathbf{k})=
\left[M_{K^*} +  \frac{m_{s} m_{\bar{q}}  +  \mathbf{k}^{2}}{z(1-z)M_{K^*}} \right]
\delta_{h,-\bar{h}} + \frac{(m_s - m_{\bar{q}})}{z(1-z)M_{K^*}} k(e^{-i \theta_k} \delta_{h+,\bar{h}+} + e^{i \theta_k} \delta_{h-,\bar{h}-})
\label{SL}
\end{equation}
and
\begin{equation}
S_{h,\bar{h}}^{K^*,T(\pm)}(z,\mathbf{k})=
\frac{\sqrt{2}}{z(1-z)} \{ [(1-z) \delta_{h\mp,\bar{h}\pm} - z
\delta_{h\pm,\bar{h}\mp} ]k e^{\pm i \theta_k} \mp [m_s-z(m_s-m_{\bar{q}})]
\delta_{h\pm,\bar{h}\pm} \} 
\label{ST}
\end{equation}
where $\mathbf{k}=ke^{i\theta_k}$ and we have again made explicit the $SU(3)$ flavor symmetry breaking correction proportional to $(m_s-m_{\bar{q}})$. Note that in the limit of exact $SU(3)$ flavor symmetry, we recover the expressions for the spinor wavefunctions of the $\rho$ meson used in Ref. \cite{Dosch:2006kz,Forshaw:2003ki,Kulzinger:1998hw} and also given in Ref.  \cite{Ahmady:2012dy}.

To twist-$3$ accuracy, four DAs parametrize the operator product expansion of meson-to-vacuum matrix elements \cite{Ball:2007zt}: 
\begin{eqnarray}
\langle 0|\bar q(0)  \gamma^\mu s(x^-)| K^*
(P,\lambda)\rangle
&=& f_{K^*} M_{K^*}
\frac{e_{\lambda} \cdot x}{P^+x^-}\, P^\mu \int_0^1 \d u \; e^{-iu P^+x^-}
\phi_{K^*}^\parallel(u,\mu)
\nonumber \\
&\hspace{-1.0cm}+&\hspace{-0.5cm} f_{K^*} M_{K^*}
\left(e_{\lambda}^\mu-P^\mu\frac{e_{\lambda} \cdot
x}{P^+x^-}\right)
\int_0^1 \d u \; e^{-iu P^+x^-} g^{\perp (v)}_{K^*}(u,\mu)  \;,
\label{DA:phiparallel-gvperp}
\end{eqnarray}
\begin{equation}
\langle 0|\bar q(0) [\gamma^\mu,\gamma^\nu] s (x^-)|K^*
(P,\lambda)\rangle =2 f_{K^*}^{\perp} (e^{\mu}_{\lambda} P^{\nu} -
e^{\nu}_{\lambda} P^{\mu}) \int_0^1 \d u \; e^{-iuP^+ x^-} \phi_{K^*}^{\perp}
(u, \mu) \label{DA:phiperp}
\end{equation}
and
\begin{equation}
\langle 0|\bar q(0) \gamma^\mu \gamma^5 s(x^-)|K^* (P,\lambda)\rangle
=-\frac{1}{4} \epsilon^{\mu}_{\nu\rho\sigma} e_{\lambda}^{\nu}
P^{\rho} x^{\sigma}  \tilde{f}_{K^*} M_{K^*} \int_0^1 \d u \; e^{-iuP^+ x^-}
g_{K^*}^{\perp (a)}(u, \mu) \label{DA:gaperp}
\end{equation}
where 
\begin{equation}
\tilde{f}_{K^*} = f_{K^*}-f_{K^*}^{\perp} \left(\frac{m_s+m_{\bar{q}}}{M_{K^*}} \right) \;.
\end{equation}
All four DAs satisfy the normalization condition
\begin{equation}
 \int_0^1 \d z \; \varphi  (z,\mu) = 1
\label{NormDA}
\end{equation}
where $\varphi=\{\phi_{K^*}^{\parallel,\perp},g_{K^*}^{\perp (v,a)}\}$ so that for a vanishing light-front distance $x^-= 0$, the definitions of the decay constants $f_{K^*}$ and  $f^{\perp}_{K^*}$ are recovered, i.e.
\begin{equation}
\langle 0|\bar q(0)  \gamma^\mu s(0)| K^*
(P,\lambda)\rangle =f_{K^*} M_{K^*} e_\lambda^{\mu}
\label{frho}
\end{equation}
and
\begin{equation}
\langle 0|\bar q(0) [\gamma^\mu,\gamma^\nu] s(0)|K^* (P,\lambda)\rangle =2 f_{K^*}^{\perp}  (e^{\mu}_{\lambda} P^{\nu} - e^{\nu}_{\lambda} P^{\mu}) \;.
\label{fperp}
\end{equation}
It follows that from Eqns. \eqref{DA:phiparallel-gvperp}, \eqref{DA:phiperp} and \eqref{DA:gaperp} that the twist-$2$ DAs are given by
\begin{equation}
f_{K^*} \phi_{K^*}^{\parallel}(z,\mu)= \int \d x^- e^{i zP^+
x^-}\langle 0|\bar q(0)
\gamma^+ s(x^-)|K^*
(P,L) \rangle \;,
\label{phiparallel}
\end{equation}
and
\begin{equation}
f_{K^*}^{\perp} \phi_{K^*}^{\perp}(z,\mu)= \frac{1} {2} \int \d x^- e^{i zP^+
x^-}
\langle
0|\bar q(0)
[e^{*}_{T\pm}.\gamma,\gamma^+] s(x^-)|K^*
(P,T(\pm)) \rangle
\label{phiperp}
\end{equation}
 while the twist-$3$ DAs are given by
\begin{equation}
f_{K^*} g_{K^*}^{\perp(v)}(z,\mu)= \frac{P^+}{M_{K^*}}\int \d x^- e^{i zP^+
x^-}
\langle
0|\bar q(0)
e^{*}_{T(\pm)}.\gamma s(x^-)|K^*
(P,T(\pm)) \rangle 
\label{gvperp}
\end{equation}
and
\begin{equation}
f_{K^*} \frac{\d g_{K^*}^{\perp (a)}}{\d z}(z,\mu)= \mp \frac{2 P^+}{M_{K^*}}\int \d x^- e^{i zP^+
x^-}
\langle
0|\bar q(0)
e^{*}_{T(\pm)}.\gamma \gamma^5 s(x^-)|K^*
(P,T(\pm)) \rangle \;.
\label{gaperp}
\end{equation}

To proceed we use the relation \cite{Forshaw:2011yj}
\begin{eqnarray}
\label{general-relation}
P^+\int \d x^- e^{ix^-zP^+} \langle 0 | \bar{q}(0)  \Gamma
s(x^-)
|K^*(P,\lambda) \rangle &=& \frac {N_c}{4\pi} \sum_{h,\bar{h}}
\int^{|\mathbf{k}| < \mu}
\frac{\d^2\mathbf{k}}{(2\pi)^2}
S^{K^*,\lambda}_{h,\bar{h}}(z,\mathbf{k}) \phi_{K^*}^{\lambda} (z,\mathbf{k}) \nonumber \\
&\times&
\left \{ \frac{\bar{v}_{\bar{h}}((1-z)P^{+},-\mathbf{k})}{\sqrt{(1-z)}}
\Gamma \frac{u_h(zP^+,\mathbf{k})}{\sqrt{z}} \right \}
\end{eqnarray}
where the renormalization scale $\mu$ appears as a cut-off on the
transverse momentum and $\Gamma$ stands for $\gamma^+$,
$[e^{*}_{T\pm}.\gamma,\gamma^+]$, $e^{*}_{T(\pm)}.\gamma$ or
$e^{*}_{T(\pm)}.\gamma \gamma^5$. The matrix element in curly
brackets can then be evaluated explicitly for each case \cite{Lepage:1980fj}: 
\begin{equation}
\frac{\bar{v}_{\bar{h}}}{\sqrt{(1-z)}}\gamma^+\frac{u_h}{\sqrt{z}} =2 P^+ \delta_{h,-\bar{h}} \;,
\label{gammap}
\end{equation}
\begin{equation}
\frac{\bar{v}_{\bar{h}}}{\sqrt{(1-z)}}
 [e^{*}_{T\pm}.\gamma,\gamma^+] \frac{u_h}{\sqrt{z}} = \mp 4\sqrt{2} P^+ \delta_{h\pm,\bar{h} \pm} \;,
\label{commutator}
\end{equation}
\begin{equation}
\frac{\bar{v}_{\bar{h}}}{\sqrt{(1-z)}}
 e^{*}_{T\pm}.\gamma \frac{u_h}{\sqrt{z}} =
\frac{\sqrt{2}}{z(1-z)} \{ [(1-z) \delta_{h\mp,\bar{h}\pm} - z \delta_{h\pm,\bar{h}\mp} ] k e^{\mp i \theta_k}  \mp [m_s - z(m_s-m_{\bar{q}})] 
\delta_{h\pm,\bar{h}\pm} \}
\label{gammaperp}
\end{equation}
and
\begin{equation}
\frac{\bar{v}_{\bar{h}}}{\sqrt{(1-z)}}
 e^{*}_{T\pm}.\gamma \gamma^5 \frac{u_h}{\sqrt{z}} =
\frac{\sqrt{2}}{z(1-z)} \{ \mp [z \delta_{h\pm,\bar{h}\mp} + (1-z)
\delta_{h\mp,\bar{h}\pm}] k e^{\mp i \theta_k} + [m_s -z (m_s-m_{\bar{q}})]
\delta_{h\pm,\bar{h}\pm} \} \;.
\label{gammaperpgamma5}
\end{equation}
We then use Eqs. \eqref{phiparallel}, \eqref{phiperp}, \eqref{gvperp} and \eqref{gaperp} in conjunction with Eqs. \eqref{gammap},  \eqref{commutator},  \eqref{gammaperp} and \eqref{gammaperpgamma5} to arrive at
\begin{equation}
\phi_{K^*}^\parallel(z,\mu) =\frac{N_c}{\pi f_{K^*} M_{K^*}} \int \d
r \mu
J_1(\mu r) [M_{K^*}^2 z(1-z) + m_f m_{s} -\nabla_r^2] \frac{\phi_{K^*}^L(r,z)}{z(1-z)} \;,
\label{phiparallel-phiL}
\end{equation}
\begin{equation}
\phi_{K^*}^\perp(z,\mu) =\frac{N_c }{\pi f_{K*}^{\perp}} \int \d
r \mu
J_1(\mu r) [m_s - z(m_s-m_{\bar{q}})] \frac{\phi_{K^*}^T(r,z)}{z(1-z)} \;,
\label{phiperp-phiT}
\end{equation}

\begin{equation}
g_{K^*}^{\perp(v)}(z,\mu)=\frac{N_c}{2 \pi f_{K^*} M_{K^*}} \int \d r \mu
J_1(\mu r)
\left[ (m_s - z(m_s-m_{\bar{q}}))^2 - (z^2+(1-z)^2) \nabla_r^2 \right] \frac{\phi_{K^*}^T(r,z)}{z^2 (1-z)^2
}
\label{gvperp-phiT}
\end{equation}
and
\begin{equation}
\frac{\d g_{K^*}^{\perp(a)}}{\d z}(z,\mu)=\frac{\sqrt{2} N_c}{\pi \tilde{f}_{K^*} M_{K^*}} \int \d r \mu
J_1(\mu r)
[(1-2z)(m_s^2- \nabla_r^2) + z^2(m_s+m_{\bar{q}})(m_s-m_{\bar{q}})]\frac{\phi_{K^*}^T(r,z)}{z^2 (1-z)^2} \;.
\label{gaperp-phiT}
\end{equation}

Note that if we assume exact $SU(3)$ flavor symmetry, i.e. if we set $m_s=m_{\bar{q}}$ in the above expressions for the four DAs, we recover, as expected, the expressions for the $\rho$ meson DAs derived in Ref. \cite{Ahmady:2012dy}.

We are also able to express the decay constants $f_{K^*}$ and $f_{K^*}^{\perp}$ in terms of the holographic AdS/QCD wavefunctions. From their definitions given by Eq. \eqref{frho} and Eq. \eqref{fperp} respectively, it follows that
\begin{equation}
\langle 0|\bar q(0)  \gamma^+ s(0)| K^*
(P,L\rangle =f_{K^*} P^+
\label{frho1}
\end{equation}
and
\begin{equation}
\langle 0|\bar q(0) [e_{T(\pm)}^* \cdot \gamma,\gamma^+] s(0)|K^* (P,T)\rangle =2 f_{K^*}^{\perp}  P^{+}  \;.
\label{frhoperp1}
\end{equation}
After expanding the left-hand-sides of Eqs. \eqref{frho1} and \eqref{frhoperp1}, we obtain
\begin{equation}
f_{K^*} M_{K^*} = \frac{N_c}{\pi}  \int_0^1 \d z
\left.[z(1-z)M^{2}_{K^*} + m_{f} m_{s} -\nabla_{r}^{2}]
\frac{\phi_{K^*}^L(r,z)}{z(1-z)}
\right|_{r=0} \;,
\label{vector-decay}
\end{equation}
and
\begin{equation}
f_{K^*}^{\perp}(\mu) =\frac{N_c}{\pi} \int_0^1 \d z (m_{s} -z(m_s-m_{\bar{q}})) \int \mu J_1(\mu r)  \frac{\phi_{K^*}^T(r,z)}{z(1-z)}
\label{tensor-decay-mu} \;.
\end{equation}

In Table \ref{tab:decay},  we compare the AdS/QCD predictions for the decay constants with the Sum Rules and lattice predictions. Note that our predictions  are obtained using constituent quark masses, i.e. $m_{\bar{q}}=0.35$ GeV for the light quark mass and $m_s=0.48$ GeV for the strange quark mass. Our prediction for the decay constant $f_K^*$ is in reasonable agreement with the experimentally measured value  shown in Table \ref{tab:decay}. The resulting AdS/QCD prediction for the ratio $f_{K*}^{\perp}/f_{K^*}$ is lower that those predicted by Sum Rules and lattice QCD at a scale of $2$ GeV. Note that our predictions for  the scale-dependent decay constant $f_{K^*}^{\perp}(\mu)$ hardly depends on  $\mu$ for $\mu \ge 1$ GeV. Our prediction for $f_{K^*}^{\perp}(\mu)$ should thus be viewed to hold at a low scale $\mu \sim 1$ GeV. 
\begin{table}[h]
\begin{center}
\[
\begin{array}
[c]{|c|c|c|c|c|}\hline
\mbox{Approach}&\mbox{Scale}~ \mu  &f_{K^*} \mbox{[MeV]} &f_ {K^*}^{\perp} (\mu) \mbox{[MeV]}&f_{K^*}^{\perp}/f_{K^*} (\mu)\\ \hline
 \mbox{AdS/QCD} &\sim 1~\mbox{GeV} & 228  &121  &0.53   \\ \hline
\mbox{Experiment}  & & 205\pm 6\footnote{From $\Gamma(\tau^- \to K^{*-} \nu_{\tau})$}  & & \\ \hline
 \mbox{SR}  & 1 ~\mbox{GeV}& 220 \pm 5&  185 \pm 10& 0.82 \pm 0.06\\ \hline
\mbox{SR}  & 2 ~\mbox{GeV}& 220 \pm 5& 162 \pm 9 & 0.73 \pm 0.04\\ \hline
\mbox{Lattice}  & 2 ~\mbox{GeV}& & &0.780 \pm 0.008 \\ \hline
\mbox{Lattice}  & 2 ~\mbox{GeV}& &  & 0.74 \pm 0.02\\ \hline
\end{array}
\]
\end{center}
\caption {Comparison between AdS/QCD predictions for the decay constants of the $K^*$ meson and Sum Rules\cite{Ball:2006eu}, lattice \cite{Becirevic:2003pn,Braun:2003jg} or experiment\cite{Beringer:1900zz}. The prediction for the decay constant $f_{K^*}^{\perp}$ is at a scale $\mu \sim 1$ GeV.}
\label{tab:decay}
\end{table}

\section{Comparison with Sum Rules DAs}

We are now in a position to compare the AdS/QCD DAs with those obtained using QCD Sum Rules. Note that Sum Rules predict the moments of the DAs: 
\begin{equation}
\langle \xi_{\parallel, \perp}^n \rangle_\mu = \int \d z \; \xi^n \phi_{K^*}^{\parallel,\perp} (z, \mu)  
\end{equation}
and that only the first two moments are available in the standard SR approach \cite{Ball:2007zt}. The twist-$2$ DA are then reconstructed as a Gegenbauer expansion
\begin{equation}
\phi_{K^*}^{\parallel,\perp}(z, \mu) = 6 z \bar z \left\{ 1 + \sum_{j=1}^{2}
a_j^{\parallel,\perp} (\mu) C_j^{3/2}(2z-1)\right\} \;. 
\label{phiperp-SR}
\end{equation}
where $C_j^{3/2}$ are the Gegenbauer polynomials and the coeffecients $a_j^{\parallel,\perp}(\mu)$ are related to the moments $\langle \xi_{\parallel,\perp}^n \rangle_\mu$ \cite{Choi:2007yu}. These moments and coefficients are determined at a low scale $\mu=1$ GeV and can then  be evolved perturbatively to higher scales \cite{Ball:2007zt}.  As $\mu \to \infty$, they vanish and the DAs take their asymptotic shapes. 
Here, we shall use here the numerical values of the hadronic parameters at $\mu=2$ GeV since the relevant hadronic scale for the $B \to K^* \gamma$ decay is $\mu=\sqrt{\Lambda_{\mbox{\tiny{QCD}}} m_b} \approx 2$ GeV. 

Similarly, explicit expressions for the twist-$3$ SR DAs are \cite{Ball:2007zt}
 
\begin{eqnarray}
 g_{K^*}^{\perp (a)}(z, \mu) & = & 6 z\bar z \frac{f_{K^*}}{\tilde{f}_{K^*}} \left\{ 1 + \left(
\frac{1}{3} a_1^\parallel (\mu) + \frac{20}{9} \kappa_{3}^\parallel (\mu) \right)
C_1^{3/2}(\xi) 
\right.\nonumber\\
&&\left.\hspace*{0pt} + \left( \frac{1}{6} a_2^\parallel (\mu) +
\frac{10}{9}\zeta_{3}^\parallel (\mu) + \frac{5}{12}\,
\omega_{3}^\parallel (\mu) -\frac{5}{24}\,
\widetilde\omega_{3}^\parallel (\mu) \!\right)
C_2^{3/2}(\xi) +
\left(\frac{1}{4}\widetilde\lambda_{3}^\parallel (\mu) - \frac{1}{8}
\lambda_{3}^\parallel (\mu) \right) C_3^{3/2}(\xi)\right\}\nonumber\\
&&{}+ 6\,
\left(\frac{m_s+m_q}{m_{K^*}} \right)\, \frac{f_{K^*}^\perp}{\tilde{f}_{K^*}} \left\{
z\bar z (2 + 3 \xi a_1^\perp + 2 (11-10z\bar z) a_2^\perp (\mu) ) + \bar
z \ln \bar z (1+3 a_1^\perp + 6 a_2^\perp (\mu) )\right.\nonumber\\
&&{}\left.\hspace*{3.1cm} + z \ln z 
(1-3 a_1^\perp (\mu) + 6 a_2^\perp (\mu))\right\}\nonumber\\
&&- 6\,
\left( \frac{m_s-m_q}{m_{K^*}} \right)\, \frac{f_{K^*}^\perp}{\tilde{f}_{K^*}} \left\{
z\bar z (9 a_1^\perp (\mu)+ 10\xi a_2^\perp (\mu)) + \bar
z \ln \bar z (1+3 a_1^\perp (\mu)+ 6 a_2^\perp (\mu))\right.\nonumber\\
&&{}\left.\hspace*{3.1cm} - z \ln z 
(1-3 a_1^\perp (\mu)+ 6 a_2^\perp (\mu))\right\}
\label{gV-SR}
\end{eqnarray}
and
\begin{eqnarray}
g_{K^*}^{\perp (v)} (z,\mu) & = & \frac{3}{4}\,(1+\xi^2) + \frac{3}{2}\,\xi^3
a_1^\parallel (\mu)+ \left\{ \frac{3}{7}\,a_2^\parallel (\mu)+ 5
\zeta_{3}^\parallel (\mu) \right\} (3\xi^2-1) + \left\{5\kappa_{3}^\parallel (\mu)
- \frac{15}{16}\,\lambda_{3}^\parallel (\mu) \right.\nonumber\\
&&{} \left. +
\frac{15}{8}\,\widetilde{\lambda}_{3}^\parallel (\mu) \right\} \xi
(5\xi^2-3)+ \left\{ \frac{9}{112}\,a_2^\parallel (\mu) +
\frac{15}{32}\,\omega_{3}^\parallel (\mu) -
\frac{15}{64}\,\widetilde\omega_{3}^\parallel (\mu) \right\} (35\xi^4-30
\xi^2+3)\nonumber\\
&&{}+\frac{3}{2}\, \left(\frac{m_s+m_q}{m_{K^*}} \right)\,\frac{f_{K^*}^\perp}{
f_{K^*}^\parallel} \left\{ 2 + 9 \xi a_1^\perp (\mu) + 2 (11-30 z \bar{z})
    a_2^\perp (\mu) \right.\nonumber\\
&&{}\left.\hspace*{3.1cm} + 
(1-3 a_1^\perp (\mu) + 6 a_2^\perp (\mu)) \ln z + (1+3 a_1^\perp (\mu) + 6 a_2^\perp (\mu)) 
\ln \bar{z} \right\}\nonumber\\
&&
    {}-\frac{3}{2}\, \left(\frac{m_s-m_q}{m_{K^*}} \right)\,\frac{f_{K^*}^\perp}{
f_{K^*}^\parallel} \left\{ 2\xi + 9 (1-2z \bar{z}) a_1^\perp (\mu) + 2 \xi (11-20 z \bar{z})
    a_2^\perp (\mu) \right.\nonumber\\
&&{}\left.\hspace*{3.1cm} + 
(1+3 a_1^\perp (\mu) + 6 a_2^\perp (\mu) ) \ln \bar{z} - (1-3 a_1^\perp (\mu)+ 6 a_2^\perp (\mu)) 
\ln z \right\}.
\label{gA-SR}
\end{eqnarray}
Notice that the higher twist-$3$ DAs depend on additionnal parameters namely $\zeta_3^{\parallel}$, $\kappa_3^{\parallel, \perp}$, $\omega_3^{\parallel, \perp}$, $\lambda_3^{\parallel, \perp}$, $\tilde{\lambda}_3^{\parallel}$, $\tilde{\omega}_3^{\parallel}$
which are determined using  Sum Rules at a scale $\mu=1$ GeV and then evolved to a scale of $\mu=2$ GeV \cite{Ball:2007zt}. For the corrections due to non-zero quark masses in Eqs. \eqref{gV-SR} and \eqref{gA-SR}, we follow \cite{Ball:2007zt} and take $m_s=0.10$ GeV and $m_q=m_s/R$ where $R=24.6$. 

\begin{figure}
\centering
\subfigure[~Twist-$2$ DA for the longitudinally polarized $K^*$ meson]{\includegraphics[width=.80\textwidth]{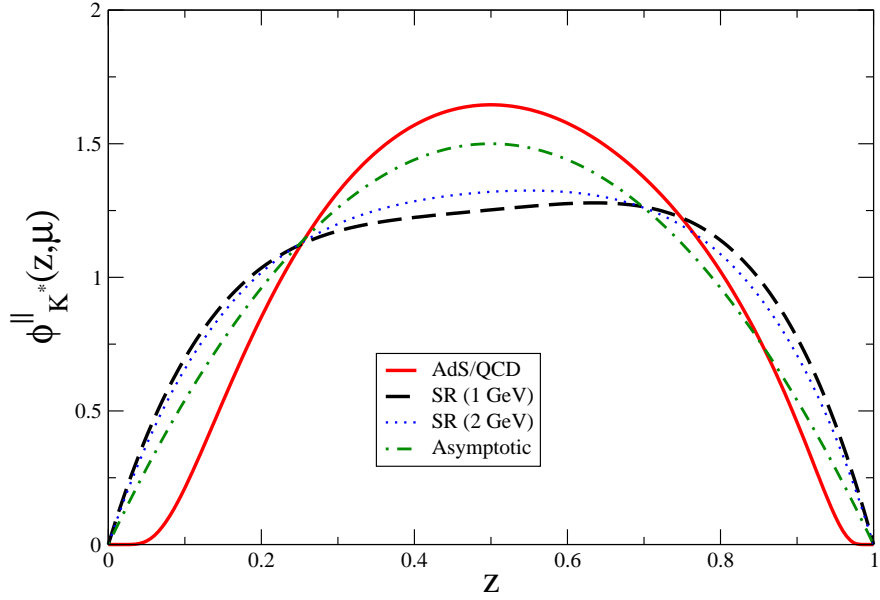} }
\subfigure[~Twist-$2$ DA for the transversely polarized $K^*$ meson]{\includegraphics[width=.80\textwidth]{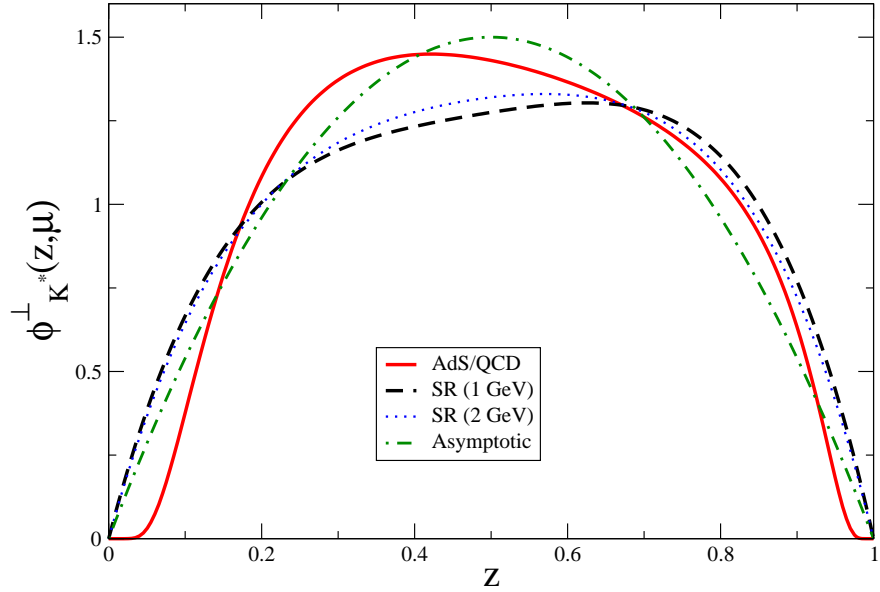} }
\caption{Solid Red: AdS/QCD DA at $\mu \sim 1$ GeV; Dashed Black: Sum Rules DA at $\mu=1$ GeV; Dotted Blue: Sum Rules DA at $\mu=2$ GeV; Dot-dashed Green: Asymptotic DA.} \label{fig:tw2DAs}
\end{figure}

\begin{figure}
\centering
\subfigure[~Derivative of the axial-vector twist-$3$ DA for the transversely polarized $K^*$ meson]{\includegraphics[width=.80\textwidth]{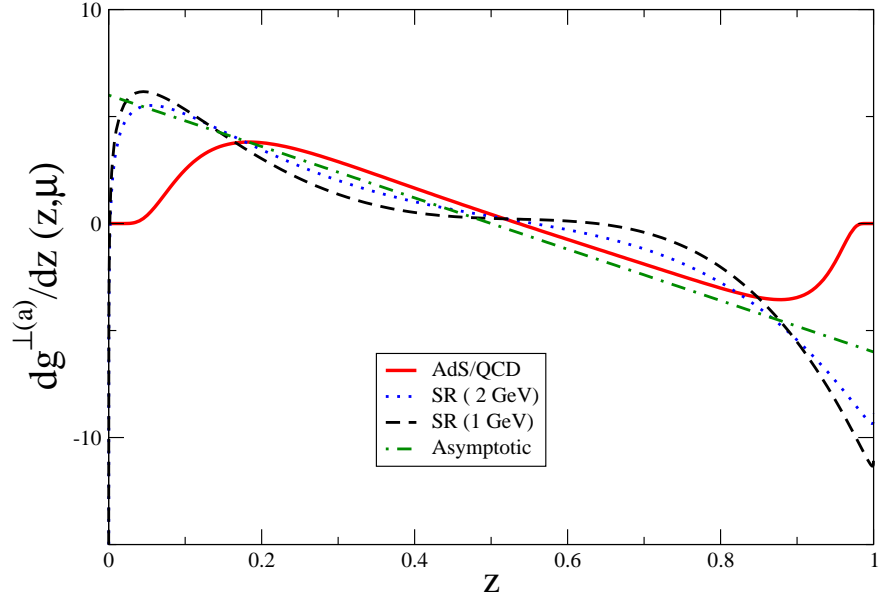} }
\subfigure[~The vector twist-$3$ DA for the transversely polarized $K^*$ meson]{\includegraphics[width=.80\textwidth]{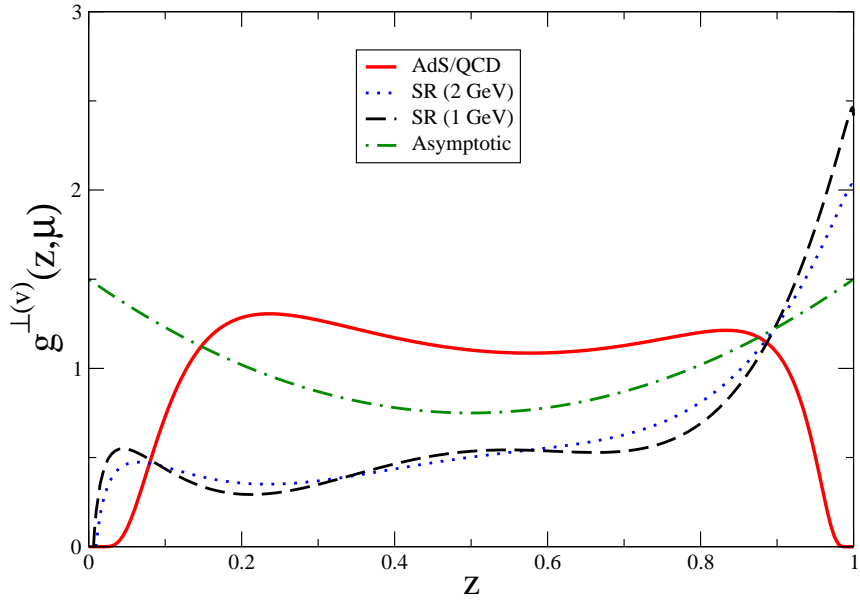} }
\caption{Solid Red: AdS/QCD DA at $\mu \sim 1$ GeV; Dashed Black: Sum Rules DA at $\mu=1$ GeV; Dotted Blue: Sum Rules DA at $\mu=2$ GeV; Dot-dashed Green: Asymptotic DA.} \label{fig:tw3DAs}
\end{figure}

In Figures \ref{fig:tw2DAs} and \ref{fig:tw3DAs}, we compare the AdS/QCD DAs at a scale $\mu \sim 1$ GeV to the Sum Rules DAs at a scale $\mu=1$ GeV and $\mu=2$ GeV and also to the asymptotic DAs.
Note that, unlike the SR DAs,  the AdS/QCD DAs lack the perturbative known evolution with the scale $\mu$ and should be viewed to be parametrizations of the meson DA at a low scale $\mu \sim 1$ GeV. The perturbative evolution of the AdS/QCD DAs can be taken into account \cite{Brodsky:2011yv} but we have not attempted to implement it here. 

We note the different end-point behaviour of the SR and AdS/QCD transverse twist-$2$ DAs. 

\section{Results and Discussions}

We can now compute the branching ratio at leading power accuracy, i.e. by using Eq. \eqref{BR}. In Table \ref{tab:integralsI1I2}, we show the SR and AdS/QCD predictions for the integrals given on the right-hand-sides of  Eq. \eqref{I1} and Eq. \eqref{I2} respectively. As can be seen the AdS/QCD and SR predictions are not very different. 
\begin{table}[h]
\begin{center}
\[
\begin{array}
[c]{|c|c|c|}\hline \mbox{Integral} & \mbox{SR} &\mbox{AdS/QCD} \\
\hline
\mbox{Eq.} \eqref{I1} & 1.99 + 2.56 i &  2.05 + 2.76 i\\ 
\hline
\mbox{Eq.} \eqref{I2} & 3.18  &  2.60\\
\hline
\end{array}
\]
\end{center}
\caption {Predictions for the integrals given on the right-hand-sides of Eq. \eqref{I1} and Eq. \eqref{I2} using the Sum Rules (SR) DA at a scale $\mu=2$ GeV and the AdS/QCD DA at a scale $\mu \sim 1$ GeV.}
\label{tab:integralsI1I2}
\end{table}
 We predict a branching ratio of $44.7 \times 10^{-6}$ using the twist-$2$ AdS/QCD DA and a branching ratio of $45.4 \times 10^{-6}$ using the corresponding SR DA. The AdS/QCD and SR prediction are therefore in agreement with each other and in agreement with the PDG values for the branching ratio quoted in Table \ref{tab:BRdata} .

We next compute the isospin asymmetry given by Eq. \eqref{isospin}.  The four convolution integrals given by Eqns. \eqref{Fperp} ,  \eqref{Gperp}, \eqref{Xperp} and \eqref{Hperp} contributing to the isospin asymmetry are given in Table \ref{tab:integrals}. As can be seen, the AdS/QCD DA does not lead to a diverging integral for $X_{\perp}$. On the other hand, the AdS/QCD results for the remaining integrals $F_{\perp}$, $G_{\perp}$ and $H_{\perp}$ are consistent with those obtained using the SR DAs. Using the AdS/QCD results, we predict an isospin asymmetry of $3.2\%$ in agreement with the most recent PDG value given in Table \ref{tab:BRdata}.

\begin{table}[h]
\begin{center}
\[
\begin{array}
[c]{|c|c|c|}\hline \mbox{Integral} & \mbox{SR} &\mbox{AdS/QCD} \\
\hline X_{\perp} & \infty &  5.65\\ 
\hline F_{\perp} & 1.14& 0.95\\
\hline G_{\perp} &2.55 + 0.43 i & 2.20  + 0.51 i\\ 
\hline H_{\perp} &2.48 + 0.50 i & 2.27 + 0.58 i\\ \hline
\end{array}
\]
\end{center}
\caption {Predictions for the four convolution integrals contributing to the isospin asymmetry using Sum Rules  DAs at a scale $\mu=2$ GeV and the AdS/QCD DAs at a scale $\mu=2$ GeV.}
\label{tab:integrals}
\end{table}

\begin{figure}
\centering
\subfigure[~The exact AdS/QCD DA compared  to its approximation by a  truncated Gegenbauer expansion.]{\includegraphics[width=.80\textwidth]{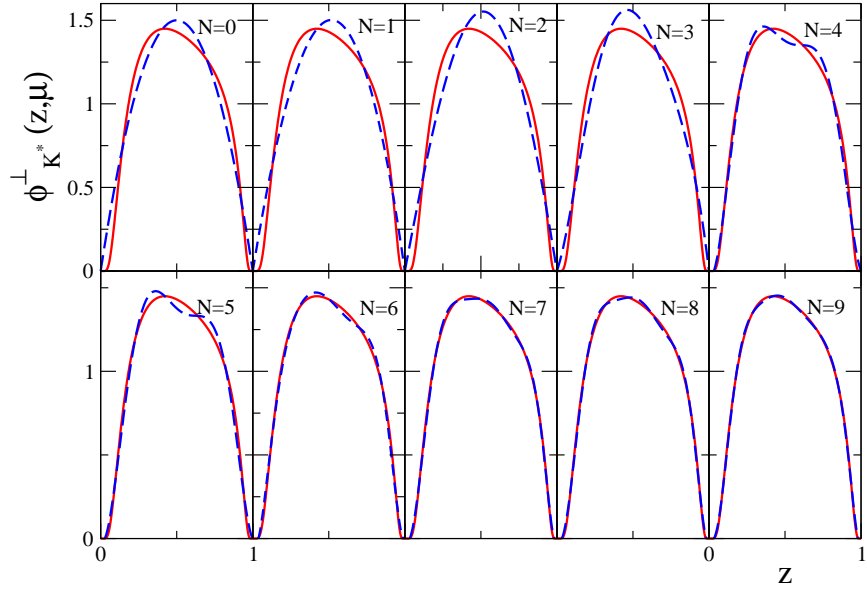} }
\subfigure[~Figure (a) in the end-point region $0.9 \le z \le 1$.]{\includegraphics[width=.80\textwidth]{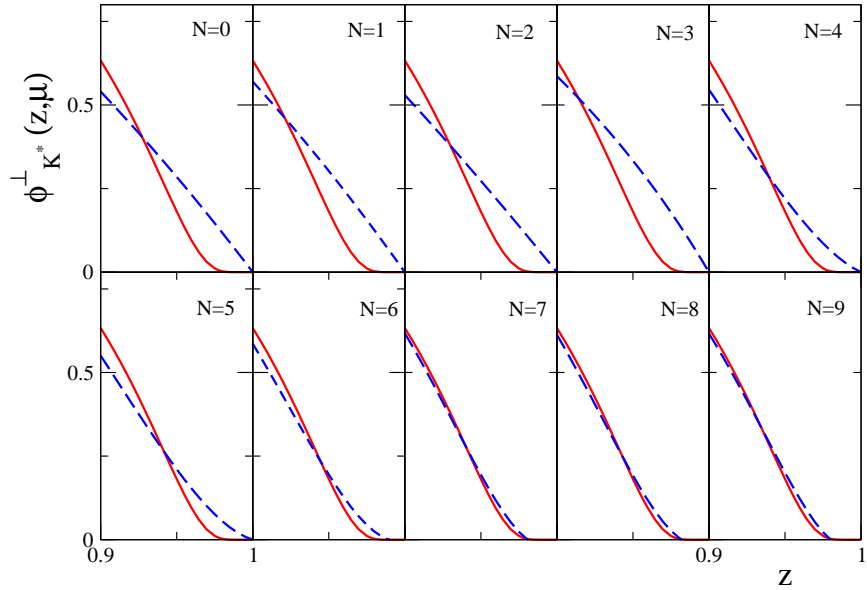} }
\caption{Solid Red: AdS/QCD twist-$2$ DA given by Eq. \eqref{phiperp}  at $\mu \sim 1$ GeV; Dashed Blue: Truncated Gegenbauer expansion given by Eq. \eqref{phiperp-AdS-Gegen}  at $\mu \sim 1$ GeV.} \label{fig:phiperpKGegen}
\end{figure}

Finally, it is instructive to investigate the origin of the end-point divergence encountered with the Sum Rules  DA but not with the AdS/QCD DA. To do so, we shall also expand the AdS/QCD DA in Gegenbauer polynomials.  We are then able to  approximate the AdS/QCD DA by a truncated Gengenbauer expansion, i.e.
\begin{equation}
\phi_N^{\perp}(z, \mu) = 6 z \bar z \left\{ 1 + \sum_{j=0}^{N}
a_j^{\perp} (\mu) C_j^{3/2}(2z-1)\right\} \;.
\label{phiperp-AdS-Gegen}
\end{equation}
where we choose $a_0^{\perp} =0$ and we determine the coeffecients $a_{j>0}^{\perp}$ by computing the moments of the AdS/QCD DA \cite{Choi:2007yu}. We then vary the number of terms $N$ from $0$ to $9$ in order to  illustrate how the truncated Gegenbauer expansion approaches the AdS/QCD DA. As  can be seen in figure \ref{fig:phiperpKGegen}, although the overall features of the AdS/QCD DA are reproduced for $N \ge 7$, the end-point behaviour is still not exactly reproduced. Moreover, it is clear that keeping only the first two terms (i.e $N=2$) in the expansion is not a good approximation to the exact AdS/QCD DA. We have checked explicitly that  the integral
\begin{equation}
X_{\perp}^{N}( \mu) = \int_0^1 \d z \; \phi^{N}_{\perp}(z, \mu) \left( \frac{1 + \bar{z}}{3 \bar{z}^2} \right)
\label{XperpN}
\end{equation}
diverges for $N \le 9$. 

We thus suspect that the end-point divergence encountered with the Sum Rules DA is due to the truncation of the Gegenbauer expansion in Eq. \eqref{phiperp-SR}. Note that this truncation is performed because Sum Rules predictions are available only for the lowest two non-vanishing moments  of the twist-$2$ DA. Thus in the Sum Rules approach, the DA is only approximately reconstructed by the Gegenbauer expansion. The higher order terms in the expansion need to be included to fully reconstruct the DA. Unless they cancel each other accidentally or the DA is evaluated at a high scale $\mu \gg \Lambda_{\mbox{\tiny{QCD}}}$, it is not a good approximation to neglect the higher order terms in the Gegenbauer expansion. In doing so, the deviation of the DA at a hadronic scale of $\mu=2$ GeV from its asymptotic form is not quantified precisely.

\section{Conclusions}

We have derived new AdS/QCD holographic DAs for the $K^*$ vector meson and we have used them in order to compute the branching ratio and isospin asymmetry for the decay $B \to K^* \gamma$. The AdS/QCD twist-$2$ DA offers the advantage of avoiding the end-point divergence encountered with the corresponding SR DA when computing the isospin asymmetry.  The resulting  prediction agrees with experiment. Moreover, the AdS/QCD prediction for the branching ratio agrees with both the SR prediction and with experiment. 

\section{Acknowledgements}
This research is supported by the Natural Sciences and Engineering Research Council of Canada (NSERC). We thank Robyn Campbell for her imput in our numerical analysis.

\bibliographystyle{apsrev}
\bibliography{Kstar_revised}

\end{document}